\documentclass[reprint,aps,prl]{revtex4-1}
\usepackage{amssymb}

\begin{document}
\title{Operator of Time and Properties of Solutions of Schr\"odinger Equation for Time Dependent Hamiltonian}
\author{Slobodan Prvanovi\'c}
\author{Du\v san Arsenovi\' c}
\affiliation{
Scientific Computing Laboratory,\\
Center for the Study of Complex Systems,\\
Institute of Physics Belgrade,\\
University of Belgrade, Pregrevica 118, 11080 Belgrade,\\
Serbia
}
\date{\today}

\begin{abstract}
Within the framework of self-adjoint operator of time in non-relativistic quantum mechanics some properties of solutions of Schr\"odinger equation, 
related to Hilbert space formalism, are investigated for two types of time dependent Hamiltonian. 
\end{abstract}

\maketitle

Ever since Schr\"odinger \cite{1}, there are many articles addressing time dependant Schr\"odinger equation and time dependent Hamiltonians \cite{2,3,4,5,6,7,8}. Within these 
investigations, one is usually concerned with the influence of environment on a system under consideration and properties of evolving states. Treatment of 
these problems entirely belongs to the standard formalism of quantum mechanics where time is treated as parameter. This means, in particular, that time 
dependence of the solutions of Schr\"odinger equation has not been systematically analyzed from the point of view of the Hilbert space formalism. Very 
often in standard quantum mechanics, time dependence of solutions of Schr\"odinger equation is seen just as exponential phase factor, so, as such, it has 
not attracted much attention. Regarding the formalism of Hilbert spaces, one is usually more concerned with the other part of solutions. That is, in the 
coordinate representation, one is interested in properties of $\psi (q)$ and not much in $e^{i f(t)}$. But, after introducing the operator of time, 
properties of the part of solution that depend on time become equally well interesting and this is what we are going to discuss in some detail.

There is a whole variety of topics and approaches related to the operator of time, {\it e.g.} \cite{9,10,11} and references therein, but let us shortly review our 
approach that we have proposed in \cite{12,13,14,15}. Our approach is similar to \cite{16}, and references therein, and \cite{17}.

In order to fulfill demand coming from general relativity, that space and time should be treated on equal footing, just like for every spatial degree of 
freedom a separate Hilbert space is introduced, one should introduce Hilbert space where operator of time $\hat t$ acts. More concretely, for the 
case of one degree of freedom, beside $\hat q$ and conjugate momentum $\hat p$, acting non-trivially in ${\cal H}_q$, there should be ${\cal H}_t$ where 
$\hat t$, together with $\hat s$ that is conjugate to $\hat t$, act non-trivially. So, in ${\cal H}_q \otimes {\cal H}_t$ for the self-adjoint operators 
$\hat q \otimes \hat I$, $\hat p \otimes \hat I$, $\hat I \otimes \hat t$ and $\hat I \otimes \hat s$ the following commutation relations hold:
\begin{eqnarray}
{1\over i\hbar} [\hat q \otimes \hat I, \hat p \otimes \hat I ] = \hat I \otimes \hat I\nonumber\\
{1\over i\hbar} [\hat I \otimes \hat t , \hat I \otimes \hat s ] = \hat I \otimes \hat I\nonumber
\end{eqnarray}
The other commutators vanish. The operator of time $\hat t$ has continuous spectrum $\{ -\infty , +\infty \}$, just like the operators of coordinate 
and momentum $\hat q$ and $\hat p$. So is the case for the operator $\hat s$, that is conjugate to time, and which is the operator of energy. After 
noticing complete similarity among coordinate and momentum on one side and time and energy on the other side, one can introduce eigenvectors of $\hat t$: 
\[
\hat t \vert t \rangle = t \vert t \rangle, \ \ \ \ \forall \ \ t \in \mathbb{R}
\]
In $\vert t \rangle$ representation, operator of energy becomes $i \hbar {\partial \over \partial t}$, while its eigenvectors $\vert E \rangle$ 
become $e^{{1\over i\hbar} Et}$, for every $E\in \mathbb{R}$.

According to Pauli, there is no self-adjoint operator of time that is conjugate to the Hamiltonian $H (\hat q, \hat p )$ which has 
bounded from below spectrum. Within our proposal, self-adjoint operator of time is conjugate to the operator of energy, which has unbounded spectrum. 
The Hamiltonian and operator of energy are acting in different Hilbert spaces, but there is subspace of the total Hilbert space where:
\[
\hat s \vert \psi \rangle = H (\hat q, \hat p ) \vert \psi \rangle  
\]
The states that satisfy this equation are physical since they have non-negative energy. The last equation is nothing else but 
the Schr\"odinger equation. By taking $\vert q \rangle \otimes \vert t \rangle$ representation of previous equation, one gets the familiar form 
of Schr\"odinger equation:
\[
i \hbar {\partial \over \partial t } \psi (q,t) = \hat H \psi (q,t) 
\]
With the shorthand notation $\hat H = \langle q \vert H (\hat q , \hat p ) \vert q' \rangle$. In other words, operator of energy has negative 
eigenvalues as well as non-negative, but the Schr\"odinger equation appears as a constraint that selects physically meaningful states. That is, 
states with non-negative energy, due to the non-negative spectra of $H (\hat q , \hat p )$, are selected by Schr\"odinger equation. For time independent 
Hamiltonian, the typical solution of Schr\"odinger equation $\psi _E (q) e^{{1\over i \hbar} Et}$ is $\vert q \rangle \otimes \vert t \rangle$ 
representation of $\vert \psi _E \rangle \otimes \vert E \rangle $, where $H (\hat q , \hat p ) \vert \psi _E \rangle = E \vert \psi _E \rangle$ 
and $\hat s \vert E \rangle = E \vert E \rangle$. The energy eigenvectors $\vert E \rangle$ have the same formal characteristics as, say, the 
momentum eigenvectors (they are normalized to $\delta (0)$ and, for different values of energy, they are mutually orthogonal). 

Now, let us discuss properties of solutions of Schr\"odinger equation with time dependent Hamiltonian. In order to be more systematic, we shall firstly 
analyze situation when time dependence appears through the term that is added to the time independent Hamiltonian, {\it i. e.}, 
$H(\hat q , \hat p) + g(\hat t)$, then the situation when the time dependent term multiplies time independent part of the Hamiltonian, {\it i.e.}, 
$H(\hat q , \hat p) \cdot g(\hat t)$, and finally we shall briefly comment combination of these two. (We are not going to discuss the case when $g(\hat t)$ 
depends on $\hat q$ and $\hat p$ since it is beyond the scope of this article.)

For the Hamiltonian $H(\hat q , \hat p) + g(\hat t)$, the Schr\"odinger equation (in $\vert q \rangle \otimes \vert t \rangle$ representation, and with 
$\hbar = 1$ taken for simplicity) is satisfied for $\psi _E (q) {1\over\sqrt{2\pi}}e^{ih(t)}e^{iEt}$, where $E$ is the eigenvalue of Hamiltonian and 
$h(t)=\int g(t) dt$. The time dependent part of the solution can be seen as modulated Dirac delta function since it is slightly different from the time 
representation of energy eigenvector, which is Dirac delta function in energy representation. The properties of modulated Dirac delta functions are as follows.

With the set of functions $\langle t\vert E,h\rangle={1\over\sqrt{2\pi}}e^{-ih(t)}e^{-iEt}$ it is easy to verify orthogonality and completeness relations:
\[
\langle E,h\vert E^\prime,h\rangle=\delta(E-E^\prime),\quad\hat I=\int_{-\infty}^{+\infty}\vert E,h\rangle\langle E,h\vert dE.
\]
The $h$ generated transform:
\[
[{\cal F}_{h}f(t)](E)\equiv{1\over\sqrt{2\pi}}\int_{-\infty}^{+\infty}f(t)e^{-ih(t)}e^{-iEt}dt
\]
is easily related to the Fourier transform:
\[
[{\cal F}_{h}f(t)](E)=[{\cal F}(f(t)e^{-ih(t)}](E).
\]
It's inverse is
\[
[{\cal F}_{h}^{-1}f(E)](t)\equiv{1\over\sqrt{2\pi}}\int_{-\infty}^{+\infty}f(E)e^{ih(t)}e^{iEt}dE.
\]
This set of functions are generalised eigenfunctions of the self-adjoint operator $\left(i{\partial\over\partial t}-g(t)\right)$:
\[
\left(i{\partial\over\partial t}-g(t)\right)\left[{1\over\sqrt{2\pi}}e^{-ih(t)}e^{-iEt}\right]=E\left[{1\over\sqrt{2\pi}}e^{-ih(t)}e^{-iEt}\right].
\]

For the Hamiltonian $H(\hat q , \hat p)\cdot g(\hat t)$, the Schr\"odinger equation (in $\vert q \rangle \otimes \vert t \rangle$ representation) is satisfied 
for $\psi _E (q) {1\over\sqrt{2\pi}}e^{-iEh(t)}$, where $E$ is the eigenvalue of Hamiltonian and $h(t)=\int g(t) dt$. The time dependent part of the 
solution can be called generalized Dirac delta function with explanation similar to the above given one. The properties of generalized Dirac delta 
functions $\langle t\vert E,h\rangle={1\over\sqrt{2\pi}}e^{-iEh(t)} \label{genfnkcone}$ are as follows. Their bi-orthogonal set is:
\[
\langle t\vert E',h,\perp\rangle={1\over\sqrt{2\pi}}h^\prime(t)e^{-iE'h(t)}
\]
since:
\begin{eqnarray}
\langle E',h,\perp\vert E,h\rangle={1\over2\pi}\int_{-\infty}^{+\infty}dt h^\prime(t)e^{-ih(t)(E-E')}=\nonumber\\
={1\over2\pi}\int_{-\infty}^{+\infty}dy e^{-iy(E-E^\prime)}=\nonumber
\delta(E-E').
\end{eqnarray}
Resolution of unity is:
\[
\hat I=\int_{-\infty}^{+\infty}dE\vert E,h,\perp\rangle\langle E,h\vert=\int_{-\infty}^{+\infty}\vert E,h\rangle\langle E,h,\perp\vert.
\]
Functions ${1\over\sqrt{2\pi}}e^{-iEh(t)}$ generate transform:
\[
[{\cal F}_hf(t)](E)\equiv{1\over\sqrt{2\pi}}\int_{-\infty}^{+\infty}f(t)e^{-iEh(t)}dt
\]
which is related to Fourier transform by:
\[
[{\cal F}_hf(t)](E)=\left[{\cal F}\left\{{dh^{-1}(t)\over dt}f(h^{-1}(t))\right\}\right](E).
\]
It is easy to verify that inverse transform is:
\[
[{\cal F}_h^{-1}f(E)](t)={1\over\sqrt{2\pi}}h^\prime(t)\int_{-\infty}^{+\infty}f(E)e^{iEh(t)}dE.
\]
The operator ${1\over h^\prime(t)}\left(i{\partial\over\partial t}\right)\label{operone} $ has ${1\over\sqrt{2\pi}}e^{-iEh(t)}$ as generalized eigenfunctions since:
\[
{1\over h^\prime(t)}\left(i{\partial\over\partial t}\right)\left[{1\over\sqrt{2\pi}}e^{-iEh(t)}\right]=E\left[{1\over\sqrt{2\pi}}e^{iEh(t)}\right].
\]
This operator is not self-adjoint, which is in accordance with the fact that ${1\over\sqrt{2\pi}}e^{-iEh(t)}$ are not orthogonal.

While:
\[
S(E)={1\over2\pi}\int_{-\infty}^{+\infty}dte^{iEh(t)}\label{gdistribution}
\]
may not be integrable, it may exist as a distribution. Acting on a Schwartz function $\varphi(E)$, it holds:
\begin{eqnarray*}
\langle S,\,\varphi\rangle=\int_{-\infty}^{+\infty}S(E)\varphi(E)dE=\\
={1\over2\pi}\int_{-\infty}^{+\infty}dt\int_{-\infty}^{+\infty}\varphi(E)e^{-iEh(t)}dE=\\
={1\over\sqrt{2\pi}}\int_{-\infty}^{+\infty}dt[{\cal F}(\varphi(E))](h(t))=\\
={1\over\sqrt{2\pi}}\int_{-\infty}^{+\infty}du{dh^{-1}(u)\over du}[{\cal F}(\varphi(E))](u).
\end{eqnarray*}
Using Parseval's theorem, the result is:
\[
\langle S,\,\varphi\rangle={1\over\sqrt{2\pi}}\int_{-\infty}^{+\infty}\varphi(E)\left\{\left[{\cal F}^{-1}\left(\left(dh^{-1}(t)\over dt\right)^*\right)\right](E)\right\}^*
\]
so:
\[
S(E)={1\over\sqrt{2\pi}}\left\{\left[{\cal F}^{-1}\left(\left(dh^{-1}(t)\over dt\right)^*\right)\right](E)\right\}^*.
\]

Finally, let us just mention that in case of $H(\hat q , \hat p)\cdot g_1 (\hat t) + g_2 (\hat t)$ one can easily combine above results. On the other side, 
for the case of $H(\hat q , \hat p)\cdot g(\hat t)$, if the whole solution of Schr\"odinger equation $\psi _E (q) {1\over\sqrt{2\pi}}e^{-iEh(t)}$ is taken into account, 
mutual orthogonality of $\psi _E (q)$ for different $E$ implies the orthogonality of the whole $\psi _E (q,t)$.

\section{Acknowledgement}
We acknowledge support of the the Serbian Ministry of education, science and technological development, contract ON171017.

\end{document}